# High field magnetoresistivity of epitaxial $La_{2-x}Sr_xCuO_4$ thin films

*J. Vanacken[1,2], L. Weckhuysen[1], T. Wambecq[1], P. Wagner[3], and V.V. Moshchalkov[1]*

[1] *Laboratorium voor Vaste-Stoffysica en Magnetisme, Katholieke Universiteit Leuven, Celestijnenlaan 200 D, B-3001 Leuven, Belgium*

[2] *Laboratoire National de Champs Magnétiques Pulsés, 143, Avenue de Rangueil BP 4245, F31432 Toulouse, France*

[3] *Instituut voor Materiaalonderzoek, Limburgs Universitair Centrum, Wetenschapspark 1, B-3590 Diepenbeek, Belgium*

**Abstract**

A large positive magnetoresistivity (up to tens of percents) is observed in both underdoped and overdoped *superconducting* $La_{2-x}Sr_xCuO_4$ epitaxial thin films at temperatures far above the superconducting critical temperature $T_c$. For the underdoped samples, this magnetoresistance far above $T_c$ cannot be described by the Kohler rule and we believe it is to be attributed to the influence of superconducting fluctuations. In the underdoped regime, the large magnetoresistance is only present when at low temperatures superconductivity occurs. The strong magnetoresistivity, which persists even at temperatures far above $T_c$, can be related to the pairs forming eventually the superconducting state below $T_c$. Our observations support the idea of a close relation between the pseudogap and the superconducting gap and provide new indications for the presence of pairs above $T_c$.

*PACS numbers*: 74.25.-q, 74.25.Dw, 74.25.Fy, 74.40.+k

**Introduction**

One of the unusual features of high $T_c$ layered superconductors is the opening of a pseudo-gap in the electronic energy spectrum at a temperature $T^*$ far above the critical temperature $T_c$. Although the existence of the pseudo-gap is commonly accepted and confirmed by several experimental techniques such as nuclear magnetic resonance (NMR), tunneling spectroscopy, angle resolved photoemission spectroscopy (ARPES) and electronic Raman scattering [1], the origin of the pseudo-gap is still not revealed. The intriguing question is whether the pseudogap and the superconducting gap have a common origin. If they are related, the pseudogap might be associated with the presence of electronic pairs above $T_c$. From this point of view, superconductivity occurs when the phase of these pairs becomes coherent and not when they are first formed in the phase incoherent state [2]. The idea of a precursory pair formation at relatively high temperatures $T_c < T < T^*$ and its relevance for high $T_c$ superconductivity is supported by the experimental observation that the pseudogap evolves into the superconducting gap at low temperatures, as clearly demonstrated by scanning tunneling spectroscopy [3]. Moreover, the ARPES data [4,5] indicate that the pseudo- and the superconducting gap both have d-wave symmetry. The fact that the $T^*(p)$- and the $T_c(p)$- lines merge in the overdoped regime [6], with p the hole density, may explain the difficulty to observe a pseudogap in the strongly overdoped case. If preformed pairs exist, they should also influence the normal state transport properties of high $T_c$ superconductors at temperatures $T_c < T < T^*$. Our paper is focussed on the in-plane magnetoresistivity of the prototype system $La_{2-x}Sr_xCuO_4$ that covers completely both the underdoped (x<0.15) and the overdoped (x>0.15) regimes. The results from earlier reports on the magnetoresistivity of this system are rather contradictory: a negative magnetoresistivity has been obtained from pulsed field transport measurements [7]; a positive [8,9,10] as well as a negative magnetoresistivity [11] has been reported from DC field measurements.

In this article, we present magnetoresistivity data of $La_{2-x}Sr_xCuO_4$ epitaxial thin films measured in pulsed magnetic fields up to 50 T and in the temperature range from room temperature down to 4.2 K. It is important to note that the magnetic field will not only destroy superconductivity but can affect, at the same time, the scattering mechanisms in the normal state. Moreover, it is not known how the new sorts of quasiparticles, introduced by theorists to explain the pseudogap phase, behave in an applied magnetic field.

Therefore, the study of the field-dependence of the resistivity is indispensable in that respect. By systematically changing the hole concentration through the variation of the Sr-content, x, we have found that superconductivity at low temperatures (T < $T_c$) and a considerable positive magnetoresistivity at high temperatures (T >> $T_c$) both appear at the same Sr-content, thus relating the large magnetoresistance with superconductivity. We will present clear evidence for precursor effects from the high field transport data.

**Experimental results**

The as-grown films were prepared by DC magnetron sputtering from stoichiometric targets [12,13]. The magnetoresistivity measurements were carried out at the pulsed field facility of the Katholieke Universiteit Leuven [14,15] by using a homemade flow-cryostat and 50 T coil. All data reported in this paper were obtained on thin films (~150 nm), patterned (1000 x 50 μm strip) for four probe measurements in the transverse geometry ($\mu_0 H \perp I$) with the magnetic field perpendicular to the film ($\mu_0 H$ // c) and the current sent along the ab-plane (I // ab).
Figures 1 to 8 present the $\rho_{ab}(\mu_0 H)$ curves measured at temperatures varying from $T >> T_c$ down to 4.2 K for the $La_{2-x}Sr_xCuO_4$ thin films with Sr content x = 0.045, 0.050, 0.055, 0.060, 0.100, 0.200, 0.250 and 0.270.

***$La_{1.955}Sr_{0.045}CuO_4$***
The $La_{1.955}Sr_{0.045}CuO_4$ sample (Figure 1) shows a very weak magnetoresistivity (less than 2 % at 45 T) in the whole temperature range. This is clearly demonstrated by the graphs A, B, C and D in the middle part of Figure 1, which present the weak, in a first approximation, quadratic magnetoresistivity at the selected temperatures 16 K, 20 K, 32 K and 176 K. No smoothing has been applied to the data, and both raising and lowering field branches are shown. The magnetoresistivity, indicated in the graphs is defined as MR = ($\rho_{ab}$(50 T) - $\rho$(0 T))/$\rho$(0 T).
The in-plane resistivity $\rho_{ab}(T)$ as a function of temperature at zero magnetic field is shown in Figure 1 at the right side of the upper frame; it serves to better orientate the magnetoresistivity measurements. The open circles denote the values of the resistivity at zero magnetic field, derived from pulsed field measurements. Since $La_{1.955}Sr_{0.045}CuO_4$ exhibits, below $T_{MI}$ ~ 100 K, a resistivity that strongly diverges when lowering the temperature (e.g. $d\rho/dT$(4.2 K) ≈ - 800 μΩcm/K), even minor heating effects in the pulsed field experiment can artificially lead to negative magnetoresistivity effects at low temperatures. We judged that the magnetoresistivity of $La_{1.955}Sr_{0.045}CuO_4$ at temperatures T > 14 K could be adequately measured up to 50 T in our setup. Indeed, no discrepancies between data taken during the rising and the lowering branch of the field pulse could be found in this temperature range, a strong indication that heating effects do not influence the results. Additional measurements in DC fields up to 8 T (not shown in this article), convincingly proved that the resistivity of the sample is only slightly magnetic field dependent below 14 K as well.

***$La_{1.95}Sr_{0.05}CuO_4$***
Upon approaching the insulator-superconductor transition in the (T,x)-phase diagram (x=0.055), a considerable positive magnetoresistivity appears (Figure 2). A magnetic field of 45 T causes an excess resistivity of 10 % in $La_{1.95}Sr_{0.05}CuO_4$ at a temperature of 10 K. With increasing temperature, the magnetoresistivity of the sample at 45 T goes down to a final decrease below 2 % around 40 K. Note that $La_{1.95}Sr_{0.05}CuO_4$ does not show a sign of superconductivity at zero magnetic field down to 1.5 K, the lowest temperature investigated. In contrast, the sample demonstrates an insulator-like behavior ($d\rho_{ab}/dT$<0) from 80.5 K ($T_{MI}$) down to the lowest temperature. For clarity, only the data taken during rising magnetic field are shown in graphs A, B, C, and D of Figure 2. The overview graph in the upper frame of Figure 2 depicts, at the different temperatures, the data at zero field (open circles) and at 45 T (solid circles).

***$La_{1.945}Sr_{0.055}CuO_4$***
Figure 3 illustrates that $La_{1.945}Sr_{0.055}CuO_4$, situated at the border of the superconducting phase, manifests strong magnetoresistivity effects. At 4.2 K, the magnetoresistivity at 45 T is 33 %; its value at 9.4 K is 18 %. Although situated very close to the insulator-superconductor transition, $La_{1.945}Sr_{0.055}CuO_4$ has a robust insulator-like behavior from 72.7 K ($T_{MI}$) down to 1.5 K at zero magnetic field, seemingly not to be correlated with the occurrence of superconductivity. The graphs A, B, C, D, E and F in the lower part of Figure 3 give a clear presentation of the evolution of the resistivity with magnetic field for the $La_{1.945}Sr_{0.055}CuO_4$ sample. At low temperatures (4.2 K), a saturating $\rho(\mu_0 H)$ behavior is observed. For intermediate temperatures (20.7 K), a quadratic magnetoresistivity at low fields evolves into a behavior that tends to saturate at higher fields. Only a quadratic behavior of the magnetoresistivity remains at sufficiently high temperatures (47.7 K). Graphs A, B, C and D only show the data taken during the increasing branch of the magnetic field pulse.

### $La_{1.94}Sr_{0.06}CuO_4$

In the $La_{1.94}Sr_{0.06}CuO_4$ compound (Figure 4), the insulating phase at low temperatures gives way to superconductivity below $T_c$ = 2.4 K. The rather low critical temperature $T_c$ = 2.4 K implies that the sample is located very close to the insulator-superconductor transition. *From figure 4 it is clear that the magnetoresistivity effects become very pronounced upon increasing the charge carrier concentration through the superconducting phase.* At 4.2 K, which is nearly two times $T_c$, a field of 45 T causes a magnetoresistivity of 330 % in $La_{1.94}Sr_{0.06}CuO_4$. Note that the resistivity is not even fully saturated at 45 T. Upon increasing the temperature, the impact of the magnetic field on the resistivity diminishes, resulting in a crossing of the $\rho_{ab}(\mu_o H)$ curves taken at temperatures below $T_{MI}$ = 63 K. The fact that the $\rho_{ab}(\mu_o H)$ curves cross each other reflects that the ground state at low temperatures, obscured below $T_c$ by the superconducting phase, has an insulating character in $La_{1.94}Sr_{0.06}CuO_4$. This observation is in agreement with the results reported previous in [7, 16] on underdoped superconducting $La_{2-x}Sr_xCuO_4$ single crystals. The temperature dependence of the resistivity at 45 T is shown at the right side of the upper frame of Figure 4 by black circles. Below the metal-to-insulator transition at $T_{MI}$ = 63 K, the resistivity at 45 T exhibits insulating properties ($d\rho_{ab}/dT<0$). The graphs A, B, C, D, E and F show the functional dependence of the resistivity versus field in detail. The magnetoresistivity tends to saturate at low temperatures (4.2 K). Similar to the $La_{1.945}Sr_{0.055}CuO_4$ compound (Figure 3), this behavior gradually evolves into a quadratic dependence (50 K) upon increasing the temperature. To a lower extent, this behavior can be as well seen in the $La_{1.95}Sr_{0.05}CuO_4$ sample presented in Figure 2.

### $La_{1.9}Sr_{0.1}CuO_4$

The magnetoresistivity in $La_{1.9}Sr_{0.1}CuO_4$ ($T_c$ = 17.5 K) is shown in Figure 5. Again, the study in high magnetic fields reveals an insulating ground state ($d\rho_{ab}/dT < 0$) behind the superconducting phase, which is hidden in zero magnetic field. The use of high magnetic fields allows us to determine the metal to insulator transition temperature $T_{MI}$ = 56 K. Below $T_{MI}$ = 56 K, the $\rho_{ab}(\mu_o H)$ curves cross each other. It looks like the curves have a single intersection in the graph at the left side of the upper frame of Figure 5. However, an enlarged view of this field region indicates that the crossing shifts systematically to higher fields when lowering the temperature. Graphs A, B, C, D, E and F in Figure 5 provide a closer look upon the field dependence of the resistivity in $La_{1.9}Sr_{0.1}CuO_4$ at the selected temperatures 4.2 K, 12 K, 18 K, 30 K, 71 K and 132.7 K. Below $T_c$ = 17.5 K, the magnetoresistivity tends to saturate. Nevertheless, a complete saturation is still absent at all temperatures. At the same time, the $\rho_{ab}(\mu_o H)$ curves do not exhibit a knee-shaped feature, marking the position of the second critical field ($H_{c2}$). Above $T_c$ = 17.5 K, the magnetoresistivity exhibits a familiar behavior: it tends to saturate at low temperatures (18 K), a quadratic dependence at low fields bends down with increasing field at intermediate temperatures (30 K) and a weak quadratic dependence remains at high temperatures (71 K). The magnetoresistivity at 45 T decreases from 30 % at 30 K down to below 2 % above 71 K. At high temperatures, for example 132.7 K, the magnetoresistivity for the superconducting $La_{1.9}Sr_{0.1}CuO_4$ sample is comparable to that of the distinctly non-superconducting $La_{1.955}Sr_{0.045}$-$CuO_4$ (Figure 1).

### $La_{1.8}Sr_{0.2}CuO_4$

In contrast to the underdoped samples, the overdoped $La_{1.8}Sr_{0.2}CuO_4$ ($T_c$ = 22.8 K) demonstrates clear knee-shaped features in its field-dependent resistivity curves $\rho_{ab}(\mu_o H)$ at temperatures $T < T_c$ (see graphs A, B and C of Figure 6). As a consequence, the superconducting transitions and the second critical field $H_{c2}(T)$ can be determined for $La_{1.8}Sr_{0.2}CuO_4$. Above $H_{c2}(T)$, the resistivity still depends on the magnetic field. At the same time, the overdoped $La_{1.8}Sr_{0.2}CuO_4$ shows magnetoresistivity, far above $T_c$, dying out with increasing temperature, similar to the superconducting underdoped samples $La_{1.94}Sr_{0.06}CuO_4$ and $La_{1.9}Sr_{0.1}CuO_4$. However, the functional dependence of the magnetoresistivity changed while crossing the threshold of optimal doping. No tendency towards saturation has been observed in the magnetoresistivity of $La_{1.8}Sr_{0.2}CuO_4$, at temperatures well above the superconducting to normal transition. This is illustrated by graphs D and E in Figure 6. A linear dependence of the resistivity with respect to the field is found at low temperatures (30.6 K). At higher temperatures (79.6 K), a quadratic behavior is predominant. $La_{1.8}Sr_{0.2}CuO_4$ exhibits a minimum in $\rho_{ab}(T)$ at $T_{MI}$ = 40 K and a crossover from metallic to insulator-like behavior upon a temperature decrease, in a high magnetic field of 45 T. A low temperature insulating behavior persisting up to *optimal doping* is reported both for $La_{2-x}Sr_xCuO_4$ single crystals [16] and for the electron-doped superconductor $Pr_{2-x}Ce_xCuO_4$ [17]. In $Bi_2Sr_{2-x}$-$La_xCuO_{6+\delta}$, it disappears at 1/8 hole doping, in the *underdoped* regime [18]. *Our results on thin films, on the other hand, show a metal to insulator-like transition in $La_{1.8}Sr_{0.2}CuO_4$, stretching well into the overdoped regime.*

### $La_{1.75}Sr_{0.25}CuO_4$

The resistivity data for the strongly overdoped $La_{1.75}Sr_{0.25}CuO_4$ compound ($T_c$ = 15.6 K) is shown in Figure 7 as a function of the magnetic field. The data for both branches of the field pulse are presented and no smoothing has been applied. The $La_{1.75}Sr_{0.25}CuO_4$ sample shows various kinds of superconducting transitions. While close to $T_c$, the $\rho_{ab}(\mu_oH)$ transition is narrow, it broadens significantly when lowering the temperature. This indicates that the irreversibility line $H_{irr}(T)$ and the second critical field $H_{c2}(T)$ gradually separate from each other. The graphs A, B and C of Figure 7 illustrate that the resistivity of $La_{1.75}Sr_{0.25}CuO_4$ is linearly depending on the field above the critical fields. Above $T_c$ (= 15.6 K), $La_{1.75}Sr_{0.25}CuO_4$ still demonstrates a considerable quadratic magnetoresistivity but the $\rho_{ab}(\mu_oH)$-curves lack any sign of saturation. Surprisingly, our high field studies disclose a metal to insulator transition around 10 K ($T_{MI}$) in $La_{1.75}Sr_{0.25}CuO_4$. Although this sample is situated deeply in the overdoped regime, the resistivity values at 50 T demonstrate a distinct upturn when lowering the temperature below $T_{MI}$. This insulator-like behavior at low temperatures could be related to a weak pseudogap feature present even in this compound, or to disorder effects.

### $La_{1.73}Sr_{0.27}CuO_4$

Finally, Figure 8 contains the field dependent in-plane resistivity data of $La_{1.73}Sr_{0.27}CuO_4$ ($T_c$ = 9.2 K), our strongest overdoped sample in this study. The superconducting transitions are clear and sharp, just like in $La_{1.75}Sr_{0.25}CuO_4$. Since $La_{1.73}Sr_{0.27}CuO_4$ has, of all our samples, the lowest normal state resistivity, the mechanical vibrations, caused by the high field-pulses, have a stronger impact on its resistivity data. The effects of the vibrations, particularly present during the lowering branch of the pulse, are clearly visible in the data taken at 12 K and 54 K, respectively shown in graphs B and D in Figure 8. The other magnetoresistivity curves, presented in Figure 8, contain only the data taken during increasing magnetic field for clarity. In contrast with the previous samples, the $\rho_{ab}(\mu_oH)$-curves of $La_{1.73}Sr_{0.27}CuO_4$ do not cross each other at fields below 45 T. Above $T_c$, the magnetic field dependences of the resistivity are essentially quadratic, thus without tendency towards saturation.

In the field and temperature range used in our experiments, the magnetoresistivity is positive for all our films, in agreement with the results of [19, 20, 21]. *Our data on thin films differ from the results of references* [7, 16] *on single crystals of* $La_{2-x}Sr_x$-$CuO_4$ *with* $x = 0.08$ *and* $x = 0.13$, *where the magnetoresistivity in the limit of high fields was found to be negative.*

## Discussion

In the analysis of superconducting fluctuations the need to *separate fluctuation and normal-state conductivity contributions* in the analysis is eminent. This is a particularly difficult task in high-$T_c$ superconductors, because both their normal state and their superconducting behavior are not yet understood. In what follows, we present a possible analysis of the magnetoresistivity data of the $La_{2-x}Sr_xCuO_4$ thin films. Our observations give new and strong evidence for stripe formation in underdoped samples.

### *Absence of magnetoresistivvity in non-superconducting underdoped $La_{1.955}Sr_{0.045}CuO_4$*

Figure 1 (lower frame) presents the in-plane magnetoresistivity data for the non-superconducting $La_{1.955}Sr_{0.045}CuO_4$ sample in a so-called Kohler-plot (($\rho-\rho_0)/\rho_0$ vs $B^2$). The figure suggests that its resistivity is proportional to $H^2$ at all temperatures. Kohler's rule [22] is however only valid when all the curves coincide. We see that the curves at 87 K and 176 K nicely overlap but that the low temperature data deviate. A violation of Kohler's rule at low temperatures can be expected for this compound since its low temperature region is characterized by variable range hopping conductivity. It is most unlikely that one can describe the charge transport in this temperature regime by a rule, based on a classical Boltzmann theory for metals.

### **Magnetoresistivity of $La_{2-x}Sr_xCuO_4$ close to the insulator- superconductor transition**

The lower frames of Figures 2 and 3, show the Kohler plots for the $La_{2-x}Sr_xCuO_4$ samples with $x = 0.05$ and 0.055, situated very close to the insulator to superconductor transition. A considerable excess magnetoresistivity, which depends not quadratically on the magnetic field but rather tends to saturate, appears at low temperatures. This contribution becomes more pronounced when increasing the charge carrier concentration through the superconducting phase, as evidenced by the Kohler plots for $La_{1.94}Sr_{0.06}CuO_4$ ($T_c$ = 2.4 K) and $La_{1.9}Sr_{0.1}CuO_4$ ($T_c$ = 17.5 K), presented in Figures 4 and 5, respectively. This evolution strongly suggests that the non-quadratic contribution to the magnetoresistivity can be attributed to *superconducting fluctuations*.

In zero magnetic field, $La_{1.95}Sr_{0.05}CuO_4$ and $La_{1.945}Sr_{0.055}CuO_4$ demonstrate an insulator-like behavior ($d\rho_{ab}/dT > 0$) at low temperatures down to at least 1.5 K. Nevertheless, the superconducting fluctuations appear up to tens of Kelvin. For the superconducting samples $La_{1.94}Sr_{0.06}CuO_4$ and $La_{1.9}Sr_{0.1}CuO_4$, the fluctuations extend over a temperature range, which exceeds several times $T_c$. For example: the $La_{1.94}Sr_{0.06}CuO_4$ compound shows at 25 K (seven times $T_c$!) a magneto-resistivity of 8 % at 50 T, which is substantially higher than the ~ 1 % for the non-superconducting $La_{1.955}Sr_{0.045}CuO_4$ sample. The fluctuations are moreover of an unusual strength: typical fields for a complete suppression of fluctuations in conventional BCS bulk superconductors should not exceed the paramagnetic limiting field $\mu_o H$ = 1.38 $T_c$. However, the $La_{1.94}Sr_{0.06}CuO_4$ system shows, at 4.2 K, a magnetoresistivity that is not even saturated at 50 T, a value which is more than a decade higher than the conventional paramagnetic limit of $\mu_o H$ = 4 T for this sample with $T_c$ ~ 2.4K.

***Influence of the normal state magnetoresistivity for (strongly) overdoped $La_{2-x}Sr_xCuO_4$.***
The lower frames of figures 6-8 present the Kohler plots for the *overdoped* samples $La_{1.8}Sr_{0.2}CuO_4$ ($T_c$ = 22.8 K), $La_{1.75}Sr_{0.25}CuO_4$ ($T_c$ = 15.6 K) and $La_{1.73}Sr_{0.27}CuO_4$ ($T_c$ = 9.2 K). We see that the part of the magnetoresistivity, which has no quadratic behaviour with respect to the applied field diminishes upon increasing the charge carrier concentration and becomes finally undetectable in the $La_{1.73}Sr_{0.27}CuO_4$ sample.

This observation can be consistently interpreted in the context of our previous results that the one-dimensional character of the charge transport (stripes) fades away upon doping [23]. As the dimensionality of the electrical transport increases from 1D to 2D (or 3D), the fluctuations should indeed become less pronounced. The evidences for stripe formation from magnetoresistivity measurements are discussed further below (see eq. (1)).

So far, we have ignored the part of the magnetoresistivity that appears in a Kohler plot as a straight line. In Figure 8, we see that magnetic field dependencies of the $La_{1.73}Sr_{0.27}CuO_4$-data are essentially $H^2$ up to 45 T. Moreover, the straight lines coincide, which implies that $H/\rho_o$ scales the magnetoresistivity and hence that *the classical Kohler's rule is for valid* in $La_{1.73}Sr_{0.27}CuO_4$. *Its magnetoresistivity can safely be attributed to the normal state.* The data for the other strongly overdoped sample, $La_{1.75}Sr_{0.25}CuO_4$, follows Kohler's rule above 30 K. For the superconducting samples with a lower Sr content (Figures 2 to 6, lower frames), the dependence with $(H/\rho_o)^2$ remains linear over a wide temperature range (far above $T_c$) but the slopes increase when lowering the temperature. The deviations are more pronounced in the underdoped samples. Many authors observed this apparent violation of Kohler's rule in high-$T_c$ systems and speculated its origin [24, 19, 20, 25, 26, 27]. In early reports, it was assumed that the violation of Kohler's rule reflected the influence of superconducting fluctuations [20, 25]. However, the observed temperature or field dependences could not be reproduced in any fluctuation theory.
At a later stage, the effect has been explained as a normal state effect, caused by the presence of two separate relaxation times in the normal state [19]. This idea relies on the resonating valence bond model, in which spin and charge are separated and are described by spinon and holon quasiparticles [28]. Within this concept the relaxation times for carrier motion normal to the Fermi surface and parallel to it are different. The former, $\tau_{tr}$, is the usual *tr*ansport relaxation time. It is related to the spinon-holon scattering, which leads to a linear $T$ dependence of the resistivity, i.e., $\tau_{tr}^{-1} \propto T$. The latter, $\tau_H$, is the transverse (*H*all) relaxation time. It is the result of the spinon-spinon scattering that varies as $T^2$ like any other fermion-fermion interaction. According to the theory, the magnetoresistivity should be proportional to the square of the Hall angle $\theta_H$. However, the experiments of Balakirev [26] and Abe [27], respectively on $La_{2-x}Sr_xCuO_4$ thin films and $La_{1.905}Ba_{0.095}CuO_4$ single crystals, convincingly proved that the latter dependence does not hold. Both authors considered the failure of Kohler's rule as an anomalous aspect of the normal-state transport in high-$T_c$ systems. According to our data, a large positive magnetoresistivity that tends to saturate at high magnetic fields, appears at low temperatures when crossing the Sr content through the insulator-superconductor transition. This tendency can be attributed to superconducting fluctuations. As follows from the data on $La_{1.73}Sr_{0.27}CuO_4$, the normal-state magnetoresistivity starts to play a considerable role at higher Sr contents.

*One dimensional character of the superconducting fluctuations in underdoped $La_{2-x}Sr_xCuO_4$*

We believe that an adequate theory to describe the magnetoresistivity in high-$T_c$ systems should account for the inhomogeneous distribution of spin-rich and charge-rich areas, i.e. with the presence of stripes (1-D charge areas) [30]. It is apparent from our data that the violation of Kohler's rule and the superconducting fluctuations are much more pronounced in pseudogapped systems, where stripes are formed. Although stripes are well established experimentally, they are, up to now, strongly neglected by theoreticians working on magnetoresistivity effects.

Figure 9 shows the magnetoconductivity $\Delta\sigma = (\sigma(0\ T) - \sigma(50\ T))$ as a function of the temperature, both presented in logarithmic coordinates for $La_{1.94}Sr_{0.06}CuO_4$ ($T_c = 2.4$ K) and $La_{1.9}Sr_{0.1}CuO_4$ ($T_c = 17.5$ K). Data were taken in the temperature range 7.7 K - 125 K and 20 K - 132 K respectively. We observe a lowering of the magnetoresistivity with temperature following a similar power-law for both samples. Surprisingly, the value of the power 1.54 is close to 3/2, which corresponds, according to the Aslamazov-Larkin (AL) [43] expression (eq. 1) to superconducting fluctuations with *a one-dimensional character*!

$$\Delta\sigma_{3D} \sim \left(\frac{T_c}{T-T_c}\right)^{1/2} \quad \Delta\sigma_{2D} \sim \left(\frac{T_c}{T-T_c}\right) \quad \Delta\sigma_{1D} \sim \left(\frac{T_c}{T-T_c}\right)^{3/2} \qquad (1)$$

The nice correspondence with the *1D* AL expression is an important result since both samples have a very different critical temperature. The dimensionality derived from the paraconductivity is moreover in good agreement with stripe models. To obtain the result of Figure 9 for $La_{1.94}Sr_{0.06}CuO_4$ and $La_{1.9}Sr_{0.1}CuO_4$, we neglected the influence of the normal state on the magnetoresistivity. Most probably, the normal-state contribution is as small as the magnetoresistivity of the heavily underdoped $La_{1.955}Sr_{0.045}CuO_4$ compound: of the order of 1 %, which justifies our procedure.

Since the contrast of the stripes with respect to their surrounding decreases upon doping, the dimensionality of the charge-transport in the overdoped samples is not well defined. Although $La_{1.8}Sr_{0.2}CuO_4$ and $La_{1.75}Sr_{0.25}CuO_4$ reveal distinct pseudogap features, they partially recover a 2D (or 3D) character. It is therefore not surprising that the magnetoconductivity of the overdoped samples could not be fitted with a simple power law with respect to the temperature. Secondly, the data for the overdoped samples are noisier because of their low resistivity, which complicates an analysis like the one presented in Figure 9. Moreover, as follows from the data on $La_{1.73}Sr_{0.27}CuO_4$, the normal-state contribution to the magnetoresistivity cannot be neglected in samples with a high doping level. At the moment, there is however no theory available, allowing an accurate evaluation of this normal-state background. *It is even not 'a priori' clear whether the violation of Kohler's rule should be attributed to fluctuations or to the normal state (or to both)*. For example, if the mobile stripes bend in a magnetic field, they may influence the normal state magnetoresistivity in an unconventional way. In this context, we would like to mention the result of Ando and coworkers [29], who have reported a possible influence on the striped structure in *non-superconducting* underdoped cuprates by a magnetic field, in the configuration where the magnetic field is applied *parallel* to the *ab*-plane. Kimura et al. [20] found a strong suppression of the magnetoresistivity in $La_{2-x}Sr_xCuO_4$ around the hole concentration with $x = 1/8$, a concentration that is related to a more static nature of the stripes [30]. Their results are at least a manifestation of the importance of the striped structure in the analysis of magnetoresistivity measurements.

*Role of the pseudogap and pre-pair electronic states*

The pseudogap is emerging as an important indicator revealing the nature of the superconductivity as well as the normal state in our high-$T_c$ samples. A possible scenario relates the pseudogap with the presence of electronic pair states far above $T_c$ [30 - 36]. The idea is that Cooper pairs are formed at a temperature $T^*$ far above $T_c$, but bulk phase coherent superconductivity is only established when long-range phase coherence is obtained below $T_c$. The models, which are based on this precursor superconductivity scenario, get growing experimental support. Scanning tunneling spectroscopy measurements clearly demonstrate that the pseudogap evolves into the superconducting gap at low temperatures [37]. Moreover, ARPES data indicate that the pseudo- and the superconducting gap both have *d*-wave symmetry [4]. Our experimental observation of a close relation between the pseudogap and the superconducting fluctuations (= precursor pairs) strongly favor these models as well. Altshuler et al. [38] questioned the interpretation of the pseudogap as the superconducting gap because a large fluctuation diamagnetism has not been observed between $T_c$ and $T^*$. Emery et al. [34] stated however that the absence of dramatic diamagnetic effects is expected if the superconducting fluctuations are *one-dimensional*, and if the Josephson coupling between stripes is small. In this case, an applied magnetic field does not cause any significant orbital motion until full phase coherence develops, close to $T_c$. To our knowledge, we are not aware of other publications, which unveiled the one-dimensional character of the superconducting fluctuations experimentally (Figure 9). We found the one-dimensional nature of the transport of precursor pairs thanks to the investigation of the magnetoresistivity very close to the insulator-superconductor transition.

The magnetoresistivity data for $La_{1.9}Sr_{0.1}CuO_4$, presented in Figure 5, do show neither clearly marked second critical fields $H_{c2}(T)$ nor saturation at high fields. Fluctuating Cooper pairs seem to exist up to very high fields, most probably above the field range accessible by our pulsed field setup. Following the ideas outlined in [34, 37], $T^*$ is the mean-field critical temperature of the superconductor rather than $T_c$. When $T^*$ is used to obtain the paramagnetic limiting field for sample $La_{1.9}Sr_{0.1}CuO_4$ ($T^* \approx 400$ K, $T_c = 17.5$ K) instead of $T_c$, a value of $\mu_o H_p \approx 700$ T is obtained, illustrating that a field of 50 T is indeed not high enough to destroy completely the preformed pairs. The ARPES research of Loeser et al. on the pseudogap in $Bi_2Sr_2CaCu_2O_{8+\delta}$ [4] revealed a binding energy of 75 meV in the precursor pairs. Thus a magnetic field of about 130 T ($\mu_o \mu_B H = k_B T$) would be needed to destroy them completely. If the idea of precursor pairs is correct, the temperature seems to be a much more critical parameter for the existence of the pairs than a magnetic field up to 50 T. The 'resistive upper critical field', as defined by a line construction, is certainly a questionable concept with respect to the underdoped high-$T_c$ compounds. It is possible that the magnetoresistivity data of the samples, which manifest a pseudogap, just reflect the behavior of the precursor pairs in a magnetic field, maybe even the localization of the pairs in a magnetic field.

Superconductivity in metals is the result of two distinct quantum phenomena, pairing and long-range phase coherence. The influence of the stripes on superconductivity is therefore two-fold. *First of all, the one-dimensional character of the charge transport favors pair formation as follows from the similarities between the pseudogap in high-$T_c$ superconductors and the spin-gap in ladder cuprates and from experiments that demonstrate a connection between the superconducting- and the pseudogap. On the other hand, the low dimensionality hinders the long-range phase coherence needed to establish bulk superconductivity. It is a well-known fact that long-range phase coherence is impossible in a purely one-dimensional system.* This is in agreement with the fact that $(La,Sr,Ca)_{14}Cu_{24}O_{41}$, the only known superconducting ladder compound, becomes superconducting under high pressure when the interactions between the ladders are enhanced.

Like already stated before, the broadening of the superconducting transitions in underdoped cuprates, both in field and temperature, is most probably due to sample inhomogeneities. However, the inhomogeneities do not reflect a bad sample quality but rather an intrinsic property, related to a low charge carrier concentration and the presence of stripes.

The superconducting transitions for overdoped and underdoped $La_{2-x}Sr_xCuO_4$ epitaxial thin films, showing up in our transport measurements in pulsed magnetic fields up to 50 T, have a completely different nature. While the transition in underdoped samples is smeared out over more than 40 T, the overdoped samples reveal well-defined second critical fields $H_{c2}$, where bulk superconductivity is suppressed. In these overdoped samples, we found an upward curvature of $H_{c2}(T)$ at low temperatures, in strong contrast with the WHH model [39, 40], which predicts a saturation of the second critical field in this temperature range. A similar anomalous behavior of $H_{c2}$ with respect to the temperature has been reported in the literature for several high-$T_c$ systems [41, 42, 29].

**Conclusion:** *Gerneric phase diagram including fluctuation area*

The pulsed field transport measurements at temperatures $T > T_c$, revealed a sudden appearance of a large positive in-plane magneto-resistivity in $La_{2-x}Sr_xCuO_4$ close to the insulator-superconductor transition at $x = 0.055$. This evolution suggests that the effect can be attributed to superconducting fluctuations. The fluctuations appear at temperatures, which exceed $T_c$ by several times. It is therefore reasonable to speak about precursory pairing far above $T_c$. By presenting the magnetoresistivity data in the form of classical Kohler-plots, we found that the superconducting fluctuations are very pronounced in *underdoped* samples. At the same time, the normal-state contribution to the magnetoresistivity dominates in *overdoped* samples. The region, where we observed superconducting fluctuations, is schematically shown in the phase diagram of Figure 10 by the shaded area. In order to evaluate this region exactly, an adequate theory is needed, *which allows to separate fluctuations and normal-state contributions to conductivity*. Unfortunately, such theory is lacking at the moment. It is however clear from our data that there is a close link between the presence of strong superconducting fluctuations and the pseudogap phase. Since fluctuations are expected to become more pronounced in systems with a reduced dimensionality, this observation is in excellent agreement with the idea that the pseudogap phase is characterized by 1D charge transport [23]. The excess conductivity of the $La_{1.94}Sr_{0.06}CuO_4$ ($T_c$ = 2.4 K) and $La_{1.9}Sr_{0.1}CuO_4$ ($T_c$ = 17.5 K) sample at 50 T indeed shows a simple power-law behavior with respect to the temperature. The experimentally found power 1.54 is close to 3/2, which is characteristic for one-dimensional fluctuations, according to the basic theory of paraconductivity, proposed by Aslamasov and Larkin [43]. Hence, our findings strongly favor stripe models [30-36] and are consistent with the idea of a precursory behavior towards superconductivity far above $T_c$.

**Acknowledgements**

The Belgian IUAP, the Flemish GOA and FWO have supported this work. J.V. is a postdoctoral fellow of the CNRS-France and the FWO – Vlaanderen.

**References**

[1] T. Timusk, B. Statt, *Rep. Prog. Phys.* **62**, 61 (1999).
[2] V.J. Emery, S.A. Kivelson, *Nature* **374**, 434 (1995).
[3] M. Kugler, Ø. Fischer, Ch. Renner, S. Ono, Y. Ando, *Phys. Rev. Lett.* **86**, 4911 (2001).
[4] Loeser96 A.G. Loeser, Z.-X. Shen, D.S. Dessau, D.S. Marshall, C.H. Park, P. Fournier, A. Kapitulnik, *Science* **273**, 325 (1996).
[5] H. Ding, T. Yokaya, J.C. Campuzano, T. Takahashi, M. Randeria, M.R. Norman, T. Mochiku, K. Kadowaki, J. Giapinzakis, *Nature* **382**, 51 (1996).
[6] P.J. White, Z.-X. Shen, C. Kim, J.M. Harris, A.G. Loeser, P. Fournier, A. Kapitulnik, *Phys. Rev. B* **54**, R15669 (1996).
[7] Ando95 Y. Ando, G.S. Boebinger, A. Passner, T. Kimura and K. Kishio, *Phys. Rev. Lett.* **75**, 4662 (1995).
[8] J.M. Harris, Y.F. Yan, P. Matl, N.P. Ong, P.W. Anderson, T. Kimura, K. Kitzawa, *Phys. Rev. Lett.* **75**, 1391 (1995).
[9] A. Lacerda, J.P. Rodriguez, M.F. Hundley, Z. Fisk, P.C. Canfield, J.D. Thompson, S.W. Cheong, *Phys. Rev. B* **49** 13, 9097 (1994).
[10] T. Kimura, S. Miyasaka, H. Takagi, K. Tamasaku, H. Eisaki, S. Uchida, K. Kitazawa, M. Hiroi, M. Sera, N. Kobayashi, *Phys. Rev. B* **53**, 8733 (1996).
[11] N.W. Preyer, M.A. Kastner, C.Y. Chen, R.J. Birgeneau, Y. Hidaka, *Phys. Rev. B* **44**, 407 (1991).
[12] B. Wuyts, Z.X. Gao, S. Libbrecht, M. Maenhoudt, E. Osquiguil, Y. Bruynseraede, *Physica C* **203**, 235 (1992).
[13] P. Wagner, K.-Q. Ruan, I. Gordon, J. Vanacken, V.V. Moshchalkov, Y. Bruynseraede, *Physica C* **356**, 107 (2001).
[14] F. Herlach, L. Van Bockstal, M. van de Burgt, G. Heremans, *Physica B* **155,** 61 (1989).
[15] F. Herlach, Ch. Agosta, R. Bogaerts, W. Boon, I. Deckers, A. De Keyser, N. Harrison, A. Lagutin, L. Li, L. Trappeniers, J. Vanacken, L. Van Bockstal, A. Van Esch, *Physica B* **216**, 161 (1996).
[16] G.S. Boebinger, Y. Ando, A. Passner, T. Kimura, M. Okuya, J. Shimoyama, K. Kishio, K. Tamasaku, N. Ichikawa, S. Uchida, *Phys. Rev. Lett.* **77**, 5417 (1996).
[17] P. Fournier, P. Mohanty, E. Maiser, S. Darzens, T. Venkateson, C.J. Lobb, G. Czjzek, R.A. Webb, R.L. Greene, *Phys. Rev. Lett.* **81**, 4720 (1998).
[18] S. Ono, Y. Ando, T. Murayama, F.F. Balakirev, J.B. Betts, G.S. Boebinger, *Phys. Rev. Lett.* **85**, 638 (2000).
[19] J.M. Harris, Y.F. Yan, P. Matl, N.P. Ong, P.W. Anderson, T. Kimura, K. Kitazawa, *Phys. Rev. Lett.* **75**, 1391 (1995).
[20] T. Kimura, S. Miyasaka, H. Takagi, K. Tamasaku, H. Eisaki, S. Uchida, K. Kitazawa, M. Hiroi, M. Sera, N. Kobayashi, *Phys. Rev. B* **53**, 8733 (1996).
[21] A. Malinovski, M.Z. Cieplak, A.S. van Steenbergen, J.A. Perenboom, K. Karpinska, M. Nberkowski, S. Guha, P. Lindenfeld, *Phys. Rev. Lett.* **79**, 495 (1997).
[22] M. Kohler, *Ann. Phys.* **32**, 211 (1938).
[23] J. Vanacken, L. Trappeniers, P. Wagner, L. Weckhuysen, V.V. Moshchalkov, Y. Bruynseraede, *Phys. Rev. B* **64**, 184425 (2001).
[24] A. Lacerda, J.P. Rodriguez, M.F. Hundley, Z. Fisk, P.C. Canfield, J.D. Thompson, S.W. Cheong, *Phys. Rev. B* **49**, 9097 (1994).
[25] K. Semba, A. Matsuda, *Phys. Rev. B* **55**, 11103 (1997) and reference therein.

[26] F.F. Balakirev, I.E. Trofimov, S. Guha, M.Z. Cieplak, P. Lindenfeld, *Phys. Rev. B* **57**, R8083 (1998).
[27] Y. Abe, Y. Ando, J. Takeya, H. Tanabe, T. Watauchi, I. Tanaka, H. Kojima, *Phys. Rev. B* **59**, 14753 (1999).
[28] P.W. Anderson, *Science* **235**, 1196 (1987).
[29] Y. Ando, G.S. Boebinger, A. Passner, L.F. Schneemeyer, T. Kimura, M. Okuya, S. Watauchi, J. Shimoyama, K. Kishio, K. Tamasaku, N. Ichikama, S. Uchida, cond-mat/9908190 (1999).
[30] J. M. Tranquada, B.J. Sternlieb, J.D. Axe, Y. Nakamura, S. Uchida, *Nature* **375**, 561 (1995) ; J.M. Tranquada, J.D. Axe, N. Ichikawa, Y. Nakamura, S. Uchida, B. Nachumi, Phys. Rev. B 54, 7489 (1996) ; J.M. Tranquada, *Physica C* **282-287**, 166 (1997); J.M. Tranquada, Phys. Rev. Lett. 78, 338 (1997); J.M. Tranquada, *Physica B* **241-243**, 745 (1998).
[31] M. Randeria, cond-mat/9710223 (1997).
[32] A. Bianconi, A. Valletta, A. Perali, N.L. Saini, *Physica C* **296**, 269 (1998).
[33] J. Maly, J. Boldizar, K. Levin, cond-mat/9805018 (1998).
[34] V.J. Emery, S.A. Kivelson, *Nature* **374**, 434 (1995); V.J. Emery, S.A. Kivelson, cond-mat/9902179 (1999); V.J. Emery, S.A. Kivelson, J.M. Tranquada, cond-mat/9907228 (1999).
[35] J. Zaanen, *Science* **286**, 251(1999).
[36] V.V. Moshchalkov, Sol. St. Comm. 86, 715 (1993); V.V. Moshchalkov, L. Trappeniers, J. Vanacken, *Europhys. Lett.* **46**, 75 (1999); V.V. Moshchalkov, J. Vanacken, L. Trappeniers, *Phys. Rev. B.* **64**, 214504 (2001).
[37] M. Kugler, Ø. Fischer, Ch. Renner, S. Ono, Y. Ando, *Phys. Rev. Lett.* **87**, 4911 (2001).
[38] B.L. Altshuler, L.B. Ioffe, A.J. Millis, *Phys. Rev. B* **53**, 415 (1996).
[39] N.R. Werthammer, E. Helfand, P.C. Hohenberg, *Phys. Rev.* **147**, 295 (1966).
[40] E. Helfand, N.R. Werthamer, *Phys. Rev.* **147**, 288 (1966).
[41] A.P. Mackenzie, S.R. Julian, A. Carrington, G.G. Lonzarich, D.J.C. Walker, J.R. Cooper, D.C. Sinclair, *Physica C* **235-240**, 233-236 (1994).
[42] M.S. Osofsky, R.J. Soulen (jr.), S.A. Wolf, J.M. Broto, H. Rakoto, J.C. Ousset, G. Coffe, S. Askenazy, P. Pari, I. Bozovic, J.N. Eckstein, G.F. Virshup, *Phys. Rev. Lett.* **71**, 2315 (1993).
[43] L.G. Aslamazov, A.I. Larkin, *Sov. Solid State* **10**, 875 (1968).

**Figure captions**

*Figure 1.* The left side of the upper frame gives an overview of the field dependence of the in-plane resistivity $\rho_{ab}(\mu_0 H)$ of $La_{1.955}Sr_{0.045}CuO_4$ at different temperatures. The right side of the upper frame depicts the temperature dependence of the in-plane resistivity $\rho_{ab}(T)$ at zero magnetic field (solid line). The open circles mark the positions where the magnetoresistivity has been measured. For the positions labeled A, B, C and D, the magnetoresistivity is shown in more detail in the middle part of the picture. The magnetoresistivity at 45 T (MR) is indicated in %. The lower frame shows the Kohler plot at temperatures T=18K, T=87K and T=176K for the 47 T pulsed field.

*Figure 2.* The left side of the upper frame gives an overview of the field dependence of the in-plane resistivity $\rho_{ab}(\mu_0 H)$ of $La_{1.950}Sr_{0.050}CuO_4$ at different temperatures. The right side of the upper frame depicts the temperature dependence of the in-plane resistivity $\rho_{ab}(T)$ at zero magnetic field (solid line). The open circles denote the values of the resistivity in zero field, derived from pulsed field measurements; filled circles mark the resistivity values at 45 T. For the positions labeled A, B, C, D, E and F, the magnetoresistivity is shown in more detail in the middle part of the picture. The magnetoresistivity at 45 T (MR) is indicated in %. The lower frame shows the Kohler plots for the $La_{1.950}Sr_{0.050}CuO_4$ sample at selected temperatures 14K < T < 72K for the 45 T pulsed field data

*Figure 3.* The magnetoresistivity data of $La_{1.945}Sr_{0.055}CuO_4$ are presented in the same way as in Figure 1. The filled circles at the right side of the upper frame mark the resistivity values at 45 T. For the positions labeled A, B, C, D, E and F, the magneto-resistivity is shown in more detail in the middle part of the picture. The magnetoresistivity at 45 T (MR) is indicated in %. The lower frame shows the Kohler plots for the $La_{1.945}Sr_{0.055}CuO_4$ sample at selected temperatures (4.2K < T < 97K) for the 47 T pulsed field data.

*Figure 4.* The left side of the upper frame gives an overview of the field dependence of the in-plane resistivity $\rho_{ab}(\mu_0 H)$ of $La_{1.94}Sr_{0.06}CuO_4$ at different temperatures. The right side of the upper frame depicts the temperature dependence of the in-plane resistivity $\rho_{ab}(T)$ at zero magnetic field (solid line). The open circles denote the values of the resistivity in zero field, derived from pulsed field measurements; filled circles mark the resistivity values at 45 T. For the positions labeled A, B, C, D, E and F, the magnetoresistivity is shown in more detail in the lower part of the picture. The magnetoresistivity at 45 T (MR) is indicated in %. The lower frame shows the Kohler plots for the $La_{1.94}Sr_{0.06}CuO_4$ sample at selected temperatures used for the 47 T pulsed field data.

*Figure 5.* The magnetoresistivity data of $La_{1.9}Sr_{0.1}CuO_4$ are presented in the same way as in Figure 1. The filled circles at the right side of the upper frame mark the resistivity values at 50 T. For the positions labelled A, B, C, D, E and F, the magnetoresistivity is shown in more detail in the middle part of the picture. The magnetoresistivity at 50 T (MR) is indicated in %. The lower frame shows the Kohler plots for the $La_{1.9}Sr_{0.1}CuO_4$ sample at selected temperatures used for the 50 T pulsed field data.

*Figure 6.* The left side of the upper frame gives an overview of the field dependence of the in-plane resistivity $\rho_{ab}(\mu_0 H)$ of $La_{1.8}Sr_{0.2}CuO_4$ at different temperatures. The right side of the upper frame depicts the temperature dependence of the in-plane resistivity $\rho_{ab}(T)$ at zero magnetic field (solid line). The open circles denote the values of the resistivity at zero field, derived from pulsed field measurements; filled circles mark the resistivity values at 45 T. For the positions labeled A, B, C, D, E and F, the magnetoresistivity is shown in more detail in the middle part of the picture. The magnetoresistivity at 45 T (MR) is indicated in % at $T > T_c$. The lower frame shows the Kohler plots for the $La_{1.8}Sr_{0.2}CuO_4$ sample at selected temperatures used for the 48 T pulsed field data.

*Figure 7.* The left side of the upper frame gives an overview of the field dependence of the in-plane resistivity $\rho_{ab}(\mu_0 H)$ of $La_{1.75}Sr_{0.25}CuO_4$ at different temperatures. The right side of the upper frame depicts the temperature dependence of the in-plane resistivity $\rho_{ab}(T)$ at zero magnetic field (solid line). The open circles denote the values of the resistivity at zero field, derived from pulsed field measurements; filled circles mark the resistivity values at 45 T. For the positions labeled A, B, C, D, E and F, the magnetoresistivity is shown in more detail in the lower part of the picture. The magnetoresistivity at 50 T (MR) is indicated in %

at $T > T_c$. The lower frame shows the Kohler plots for the $La_{1.75}Sr_{0.25}CuO_4$ sample at selected temperatures used for the 50 T pulsed field data.

*Figure 8.* The left side of the upper frame gives an overview of the field dependence of the in-plane resistivity $\rho_{ab}(\mu_0 H)$ of $La_{1.73}Sr_{0.27}CuO_4$ at different temperatures. The right hand side of the upper frame depicts the temperature dependence of the in-plane resistivity $\rho_{ab}(T)$ at zero magnetic field (solid line). The open circles denote the values of the resistivity at zero field, derived from pulsed field measurements; filled circles mark the resistivity values at 45 T. For the positions labeled A, B, C and D, the magnetoresistivity is shown in more detail in the middle part of the picture. The lower frame shows the Kohler plots for the $La_{1.73}Sr_{0.27}CuO_4$ sample at selected temperatures used for the 45 T pulsed field data (left side) and the 12 T pulsed field data (right side).

*Figure 9.* The logarithm of the magnetoconductivity ($\sigma(0\,T) - \sigma(50\,T)$) as a function of the logarithm of the temperature, which is rescaled with respect to $T_c$. The constant $C = 1\,\mu\Omega cm$ accounts for the units along the y-axis. The linear fits follow the equations y = -5.5 + 1.55 x and y = -7.7 + 1.54 x for $La_{1.94}Sr_{0.06}CuO_4$ and $La_{1.9}Sr_{0.1}CuO_4$ respectively.

*Figure 10.* (T,x)-phase diagram of the $La_{2-x}Sr_xCuO_4$ samples. The antiferromagnetic (AF) and superconducting (SC) regions are indicated. The crossover temperature $T^*$ separates region I from region II, where $La_{2-x}Sr_xCuO_4$ has a pseudogap. $T_{MI}$ marks the metal to insulator transition and defines region III. The dimensions of the electronic transport in region I and II, are labelled in the Figure. A shaded area is added, which schematically shows the region, where we observed superconducting fluctuations.

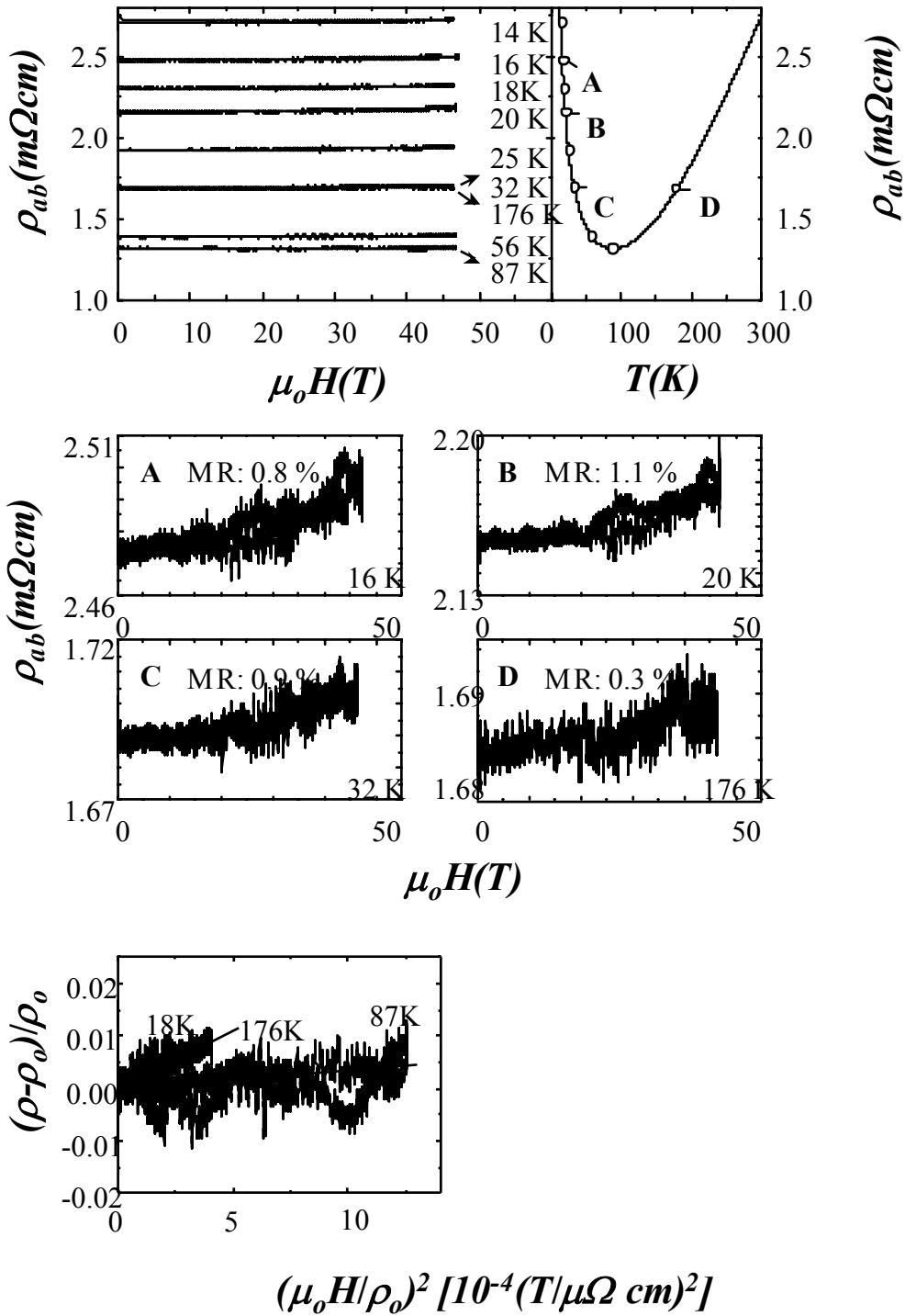

Fig.01

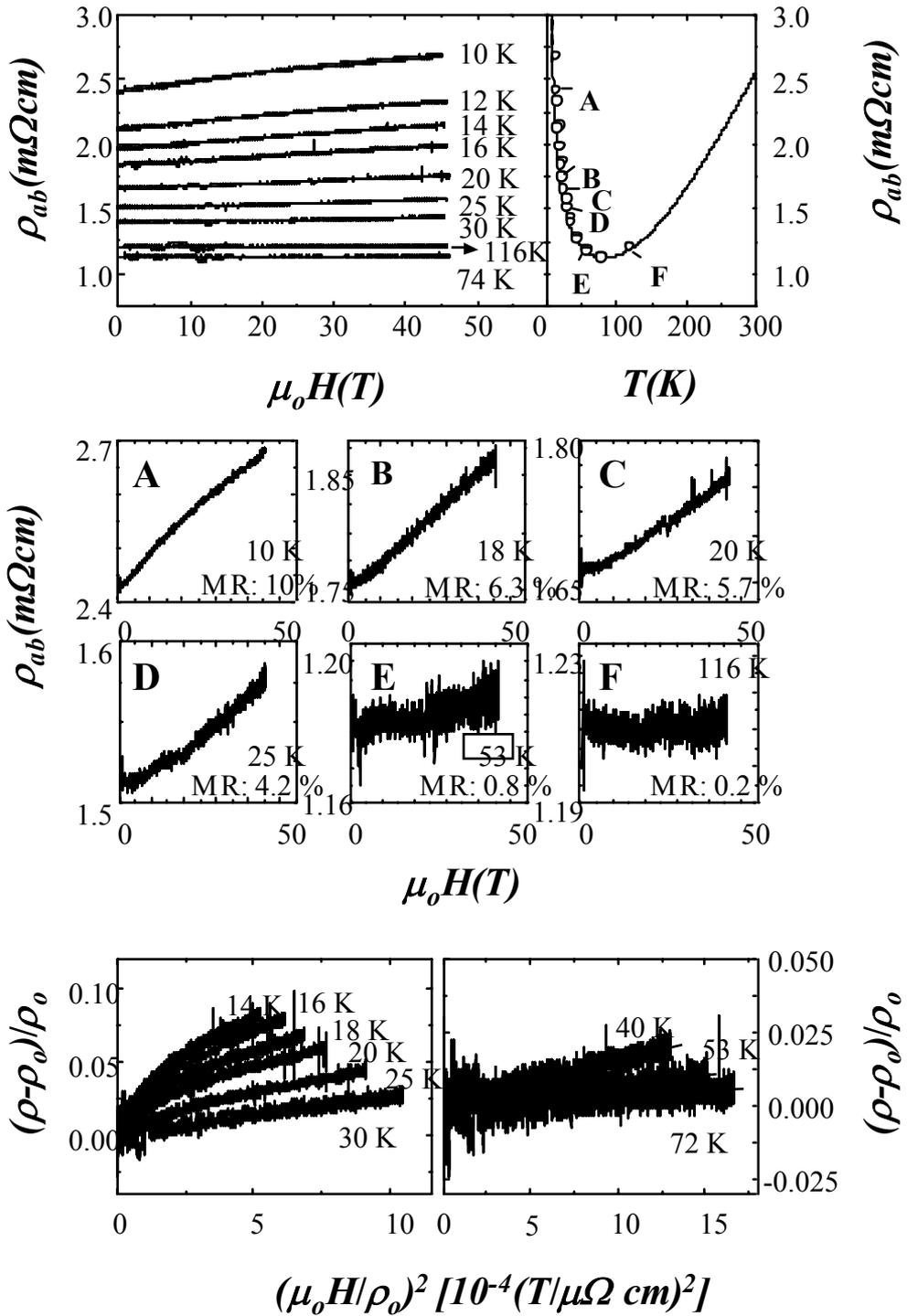

Fig.02

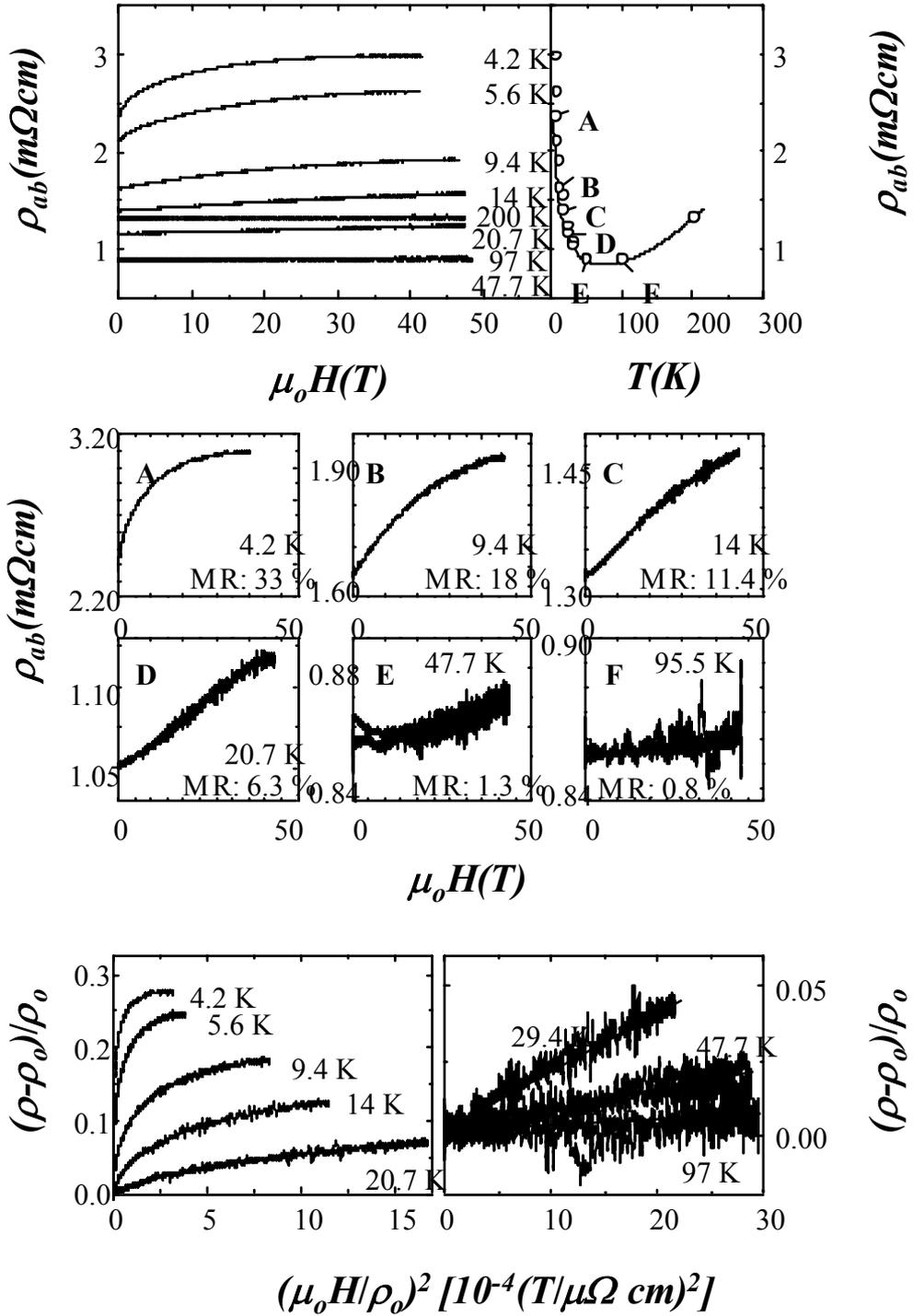

Fig.03

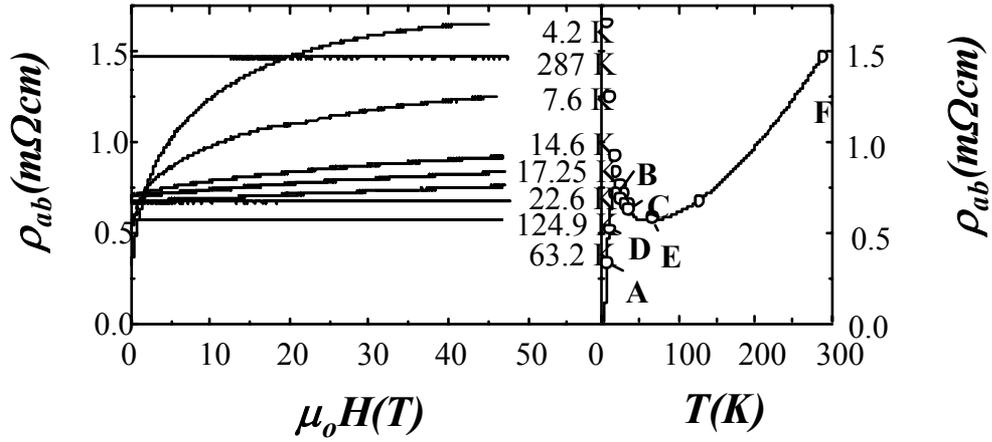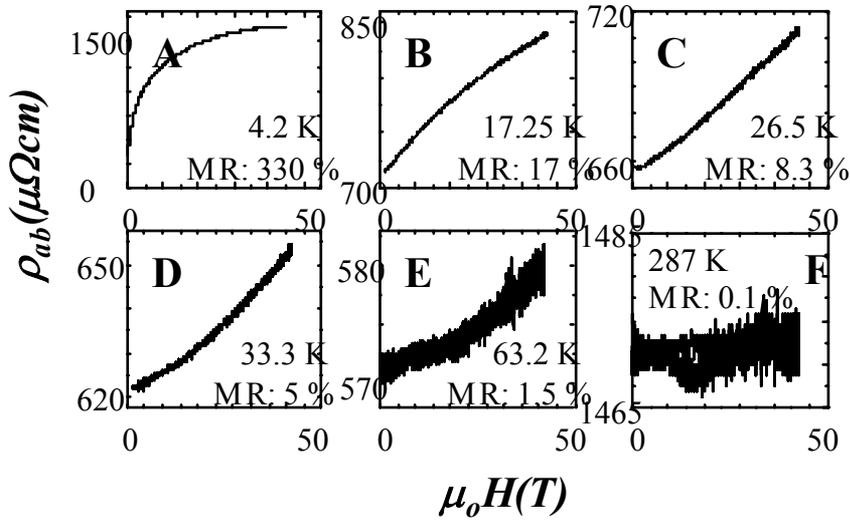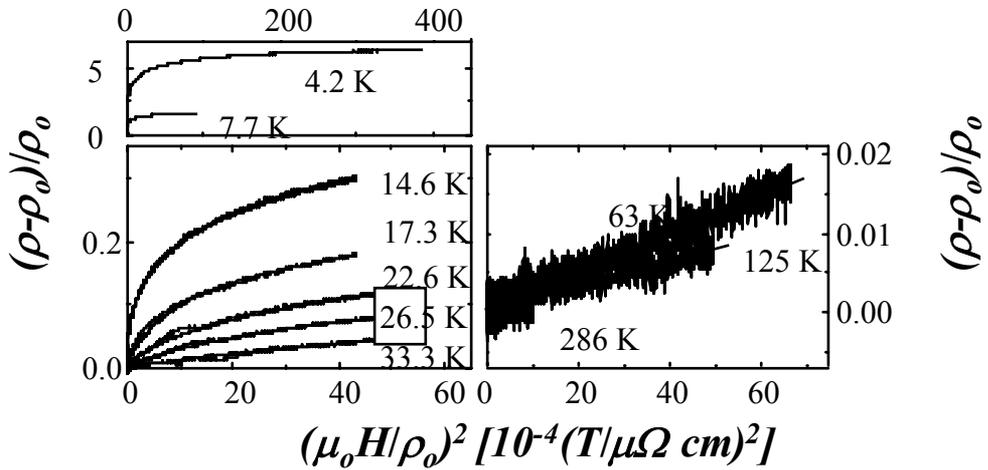

Fig.04

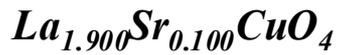
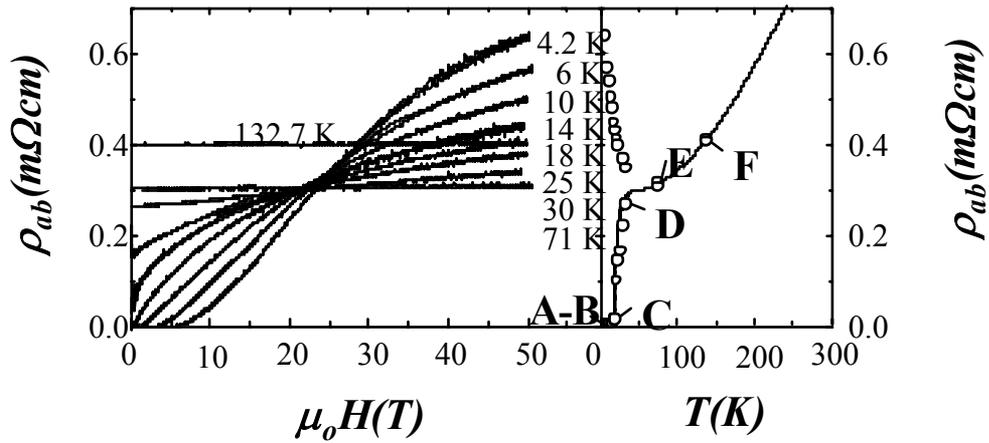
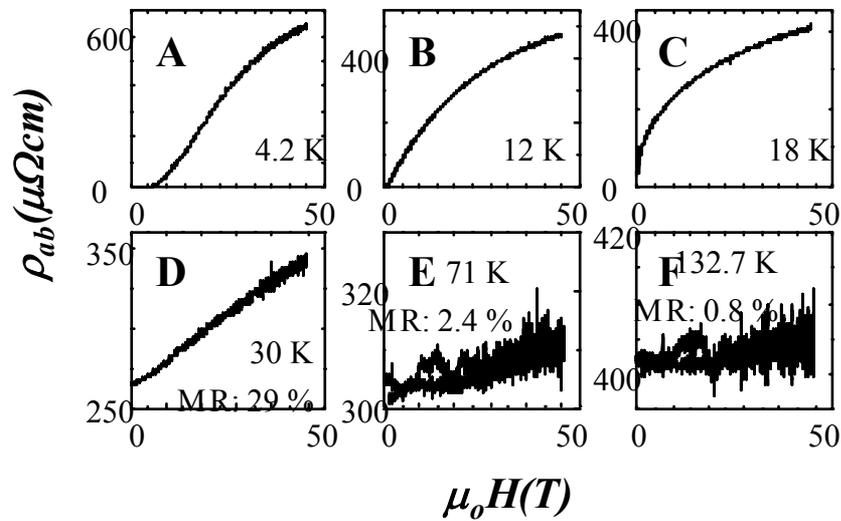
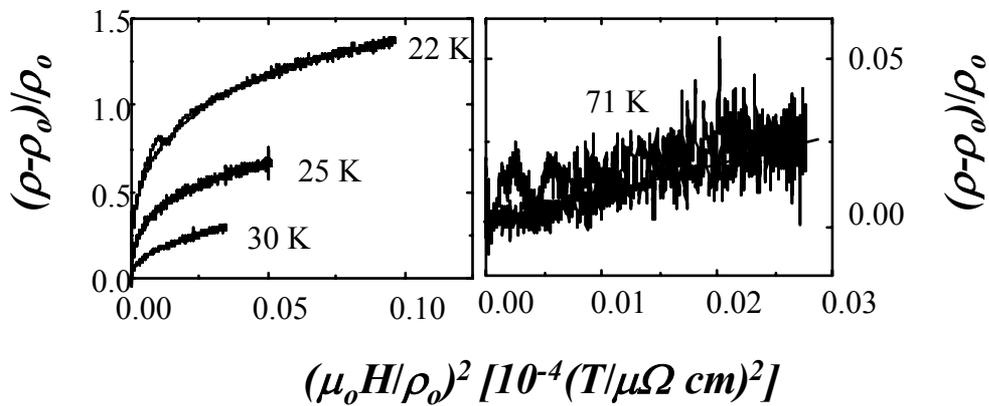

Fig.05

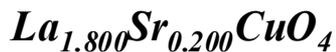
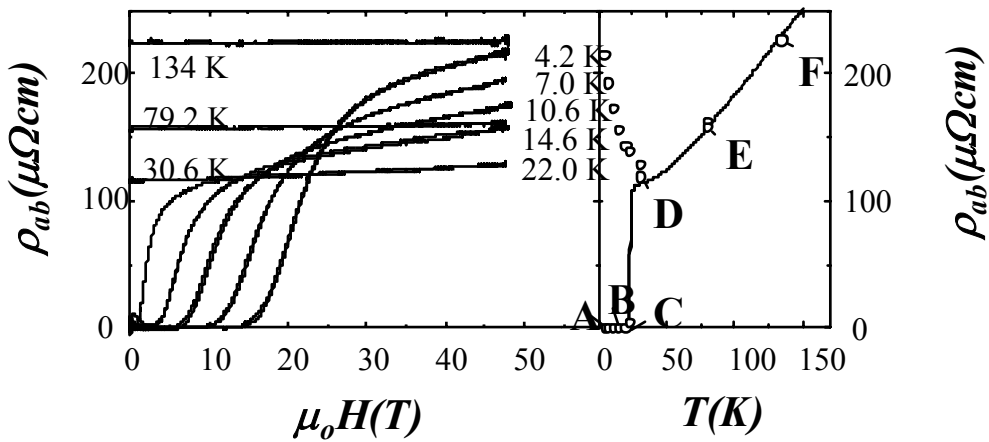
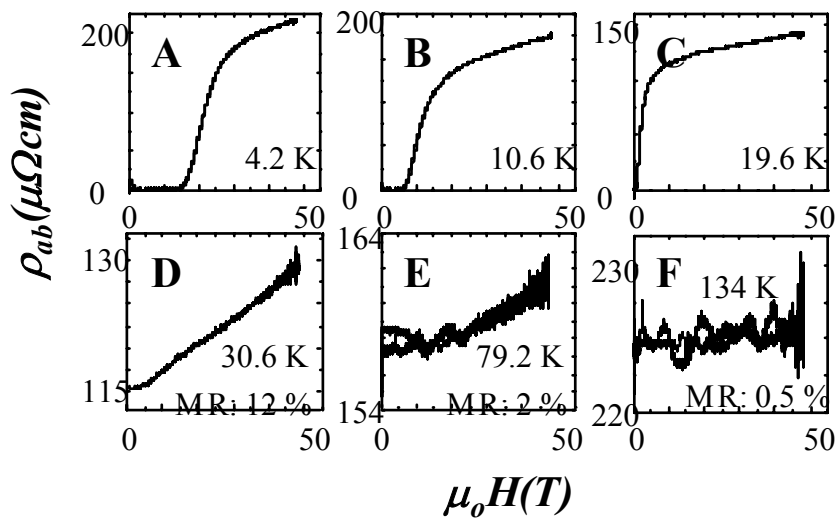
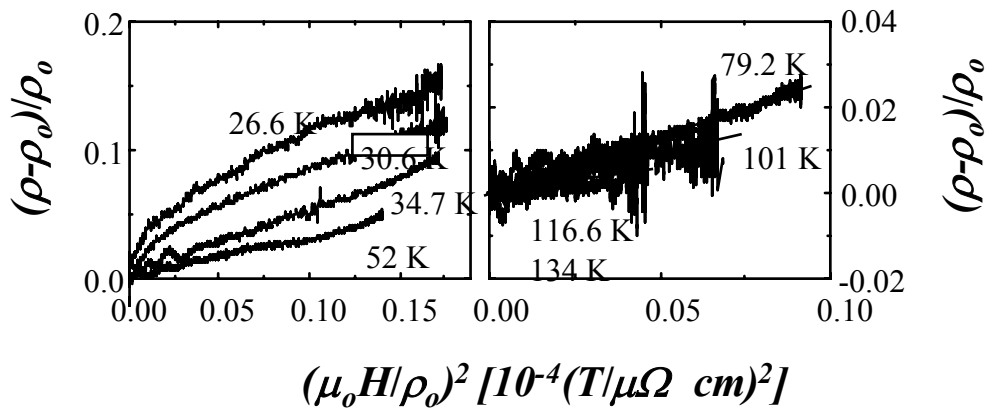

Fig.06

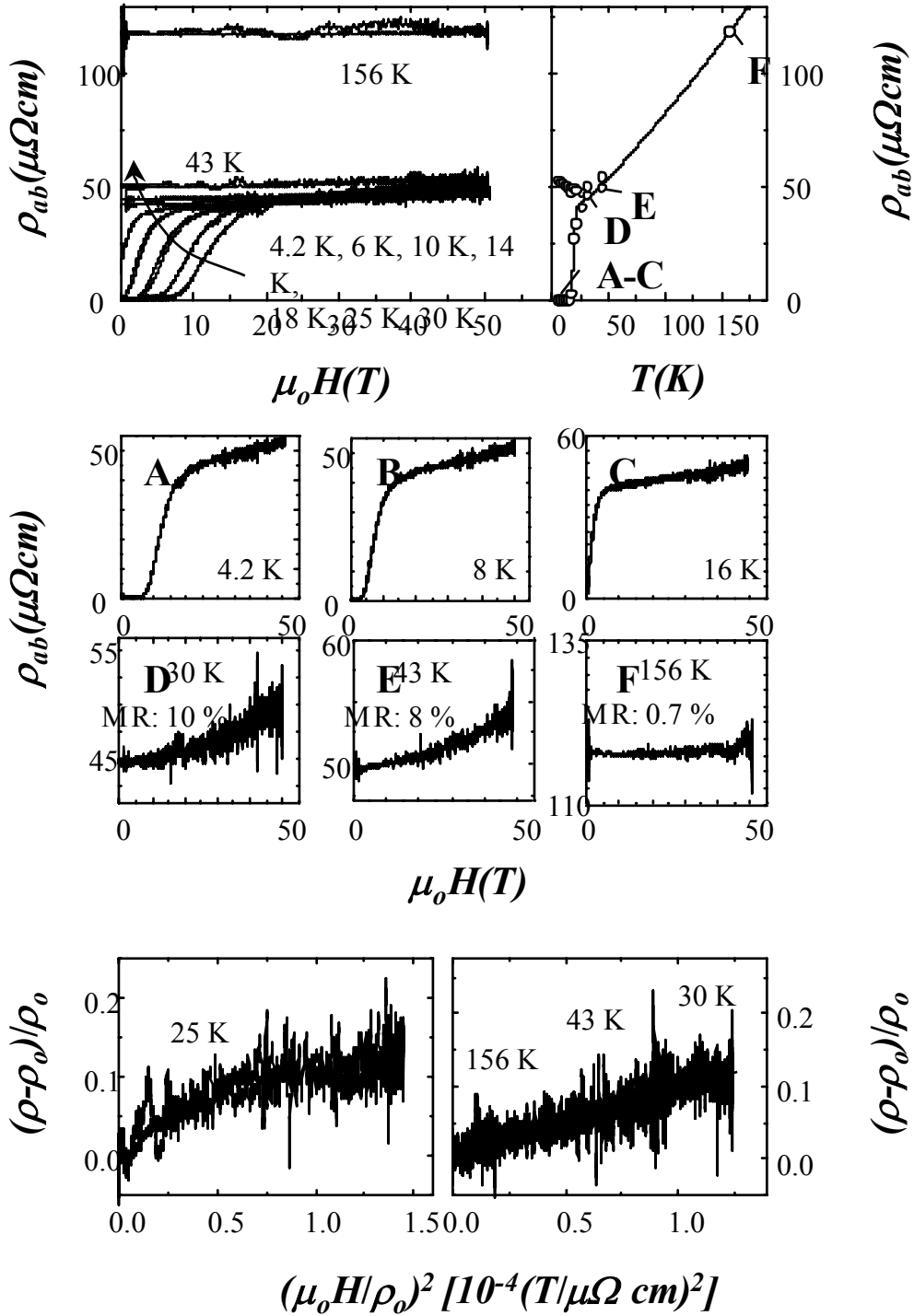

Fig.07

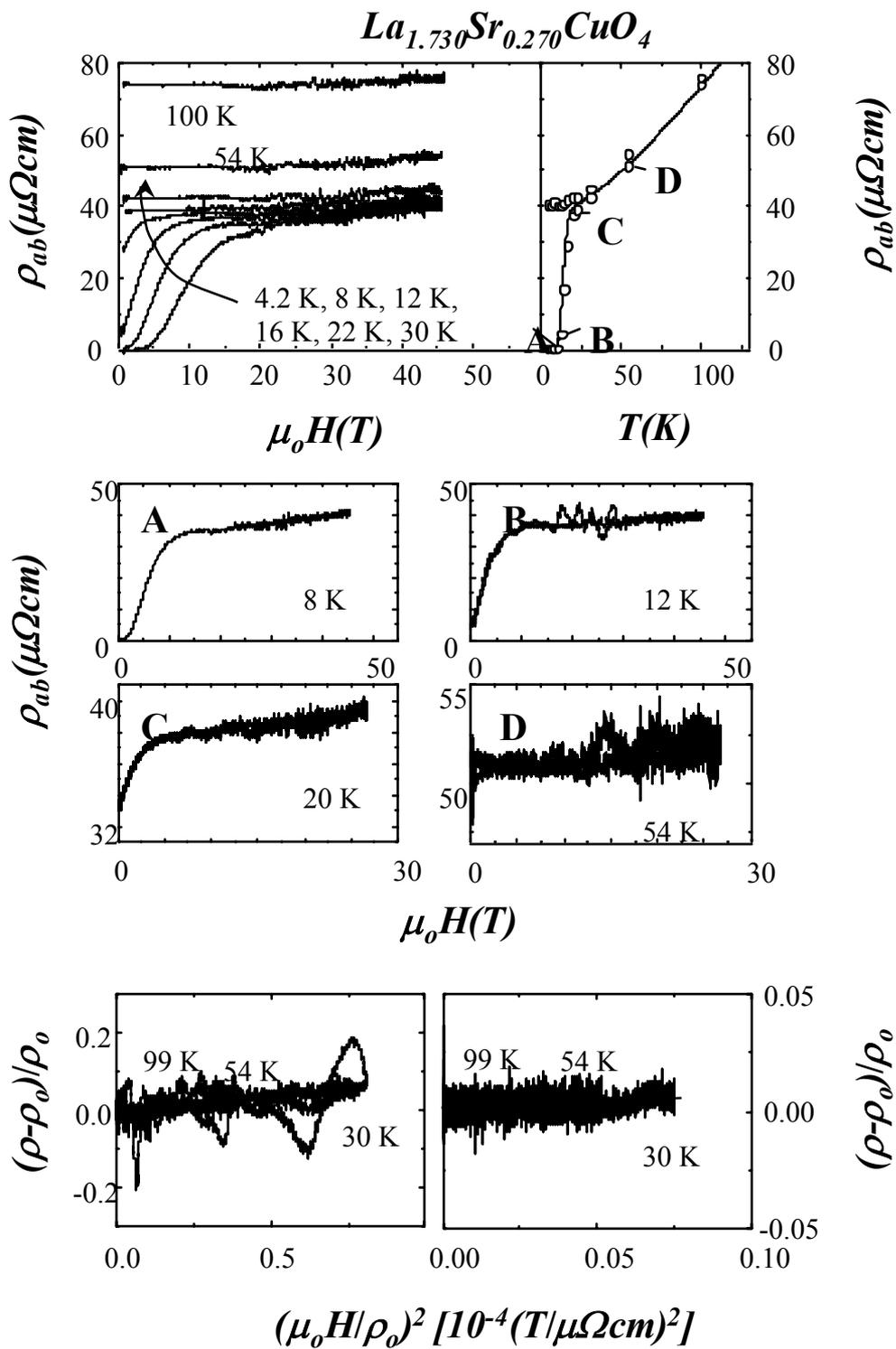

Fig.08

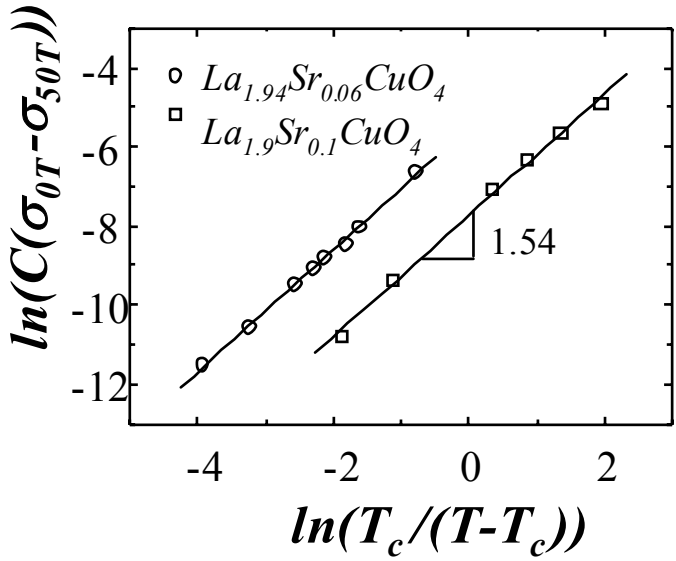

Fig.09

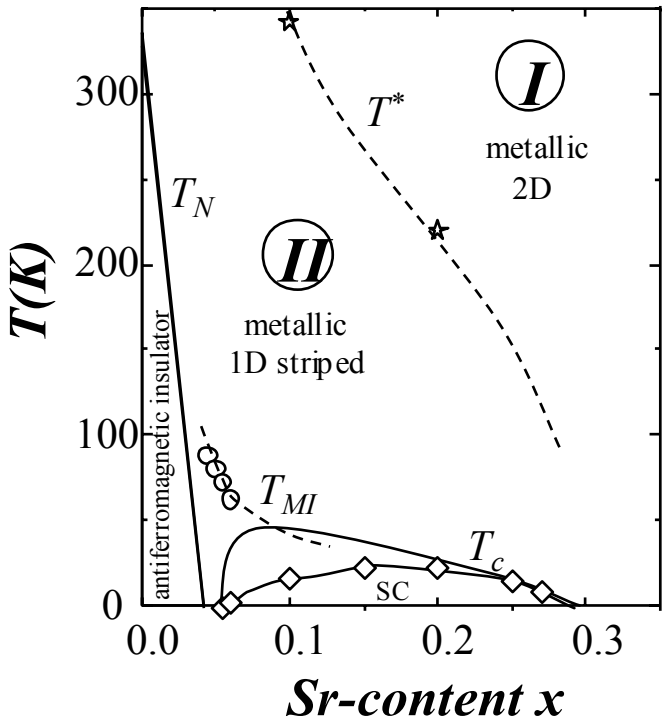

Fig.10